\documentclass[aps,pra,twocolumn,superscriptaddress,showpacs,showkeys,amsmath,amssymb]{revtex4}

\usepackage{amsfonts}
\usepackage{amssymb,amsmath}
\usepackage{mathrsfs}
\usepackage{latexsym}
\usepackage{amsmath}
\usepackage[cp1251]{inputenc}
\usepackage{graphicx}
\usepackage{dcolumn}
\usepackage{bm}
\usepackage{color}

\RequirePackage{ifthen}
\RequirePackage[pdfstartview=FitH]{hyperref}
\begin{document}

    \title{Static impurities in a weakly-interacting Bose gas}

\author{G.~Panochko} \affiliation{Department of Optoelectronics and Information Technologies, Ivan Franko National University of Lviv, 107 Tarnavskyj Str., Lviv, Ukraine}
\author{V.~Pastukhov\footnote{e-mail: volodyapastukhov@gmail.com}} \affiliation{Professor Ivan Vakarchuk Department for Theoretical Physics, Ivan Franko National University of Lviv, 12 Drahomanov Str., Lviv, Ukraine}

    \date{\today}



    \begin{abstract}
    We present a comprehensive discussion of the ground-state properties of dilute $D$-dimensional Bose gas interacting with a few static impurities. Assuming the short-ranged character of the boson-impurity interaction, we calculate the energy of three- and two-dimensional Bose systems with one and two impurities immersed.
    \end{abstract}

    \maketitle

\section{Introduction}
\label{sec1}
\setcounter{equation}{0}
The problem of impurities in mediums formed by bosons is comprehensively studied in the condensed matter physics. Even properties of a single atom immersed in the weakly-interacting Bose gas change drastically \cite{Tempere2009,Vlietinck2015,Ardila2015,Grusdt2015,Vakarchuk2017}. Depending on the strength of the boson-impurity interaction, a number of physically distinct impurity phases can be realized, namely, the Bose-polaronic state \cite{Astrakharchik2004,Novikov2009,Christensen2015,Vakarchuk2018} in various spatial dimensions, which is very similar to the free-particle one but with the modified, due to the presence of bath, kinematic characteristics; the molecular state \cite{Rath2013,Li2014}, when the impurity captures one boson with the formation of a dimer; a set of the Efimov states \cite{Levinsen2014,Levinsen2015_2,Naidon2017,Sun} with the universal scaling behavior of energy levels, and higher-order conglomerates \cite{Wang,Casteels2013,Blume2014,Shi2018,Yoshida2018,Blume2019} which involve a larger number of host atoms. Remarkably, some of these phases can be observed in experiments \cite{Jorgensen,Hu}. The experimental progress in the field of ultra-cold atomic gases has recently lead to the observation \cite{Yan} of Bose polarons at finite temperatures. This experiment confirmed previous theoretical predictions \cite{Levinsen_temp,Guenther2018,Pastukhov2018,Liu2019,Field2020,Pascual2021} about the breakdown of the quasi-particle picture description of Bose polarons in a close vicinity of the Bose-Einstein condensation (BEC) point.

Recently, the problem of two impurities immersed in the dilute one- and three-dimensional Bose gases has become a subject of extensive examination. Physically this problem is substantially distinguishable from the single Bose polaron one due to the emergence of the induced effective interaction \cite{Zinner2013,Zinner2014,Ardila2018,Guardian2018} between impurity particles. In 1D, the character of this interaction crucially depends on a sign of the boson-impurity coupling constant \cite{Brauneis2021}, the effective attraction is found for positive couplings, while the induced repulsive potential is inherent for the negative interactions. While it increases, the induced attractive interaction between impurities leads to the formation of bipolarons \cite{Petkovich} in continuum and on the lattice \cite{Pasek2019}, and even to emergence of the two-polaron bound states \cite{Will2021}. In one-dimensional geometries with harmonic trapping, the induced interaction causes the clustering \cite{Dehkharghani2018} of two initially non-interacting atoms, and modifies their quench dynamics \cite{Mistakidis2020}. By switching the boson-impurity interaction in 3D dilute BEC with two impurities, the transition from a weakly-interacting through the Yukawa potential bipolarons to the Efimov trimer state was predicted in Ref.~\cite{Naidon}. Recently, properties of a single polaron in 2D BEC have been discussed both analytically \cite{Pastukhov2018_2} and numerically \cite{Astrakharchik2D, Akaturk}. The arbitrary $D$ one-polaron case was considered in Ref.~\cite{Khan}. As far as we know the problem of two Bose polarons in 2D Bose gas has been never discussed, therefore the objective of this paper is to make the first step toward the revealing of peculiarities of the bipolaron physics and boson-induced effective interaction between impurities by considering the static limit. The absence of the impurity dynamics in this limit allows to find the exact solution of the problem in the dilute 1D Bose mediums both in one- \cite{Kain2018} and two-particle \cite{Reichert2019,Reichert2019_2} cases. In 3D, only a case of the ideal Bose gas \cite{Panochko2021,Drescher2021} is the exactly tractable one, while the presence of a weak boson-boson interaction requires \cite{Levinsen2021} a substantial numerical efforts.

\section{Formulation}
\subsection{Model}
The discussed model consists of the $D$-dimensional (here we focus on $D=2, 3$ cases) Bose gas loaded in volume $L^D$ (with the periodic boundary conditions imposed) with the weak interparticle interaction and microscopic number $\mathcal{N}$ of heavy (infinite-mass) impurities immersed in it. Heavy particles are supposed to be randomly placed in positions $\{{\bf r}_j\}$. In the following, we adopt the imaginary-time path-integral approach with Euclidean action
\begin{eqnarray}\label{S}
S=\int d x \psi^*(x)\left\{\partial_{\tau}-\varepsilon+\mu-\Phi({\bf r})\right\}\psi(x)\nonumber\\
-\frac{g_{B,\Lambda}}{2}\int d x|\psi(x)|^4,
\end{eqnarray}
where $x=(\tau, {\bf r})$ denotes the `position` in $D+1$-dimensional space (and consequently $\int d x=\int_0^{\beta}d\tau\int_{L^D}d{\bf r}$), complex field $\psi(x)$ is periodic in $\tau$ with period $\beta$ (which is the inverse temperature of the system). We also use the shorthand notations for bosonic dispersion $\varepsilon=-\frac{\hbar^2\nabla^2}{2 m}$ and the chemical potential $\mu$ that fixes average density $n$ of Bose gas, and for term
\begin{eqnarray}\label{Phi}
\Phi({\bf r})=\sum_{1\le j\le \mathcal{N}}g_{I,\Lambda}\delta_{\Lambda}({\bf r}-{\bf r}_j),
\end{eqnarray}
that describes interaction between Bose particles and impurities. The $\delta$-like two-body potential is ill-defined in the higher ($D\ge 2$) dimensions, and therefore, in order to obtain any reasonable results one should adopt some renormalization scheme. The latter is typically realized by the implication of the ultraviolet cutoff $\Lambda$ in all momentum summations and in the simultaneous rewriting of bare couplings $g_{B,\Lambda}$ and $g_{I,\Lambda}$ via the two-body vacuum binding energies $\epsilon_B$ and $\epsilon_I$
\begin{eqnarray}
g^{-1}_{B,\Lambda}=g^{-1}_B-\frac{1}{L^D}\sum_{{\bf k}}\frac{1}{2\varepsilon_k},\label{g_bare_B}\\
g^{-1}_{I,\Lambda}=g^{-1}_I-\frac{1}{L^D}\sum_{{\bf k}}\frac{1}{\varepsilon_k},\label{g_bare_I}
\end{eqnarray}
respectively (from now on we assume that all summations over the wave-vector ${\bf k}$ are restricted from the above $|{\bf k}|<\Lambda$). Such a `regularization' is already used in the definition of the point-like boson-impurity interaction potential, $\delta_{\Lambda}({\bf r})=\frac{1}{L^D}\sum_{|{\bf k}|<\Lambda}e^{{\rm i}{\bf k}{\bf r}}$, in Eq.~(\ref{Phi}). The `observable' couplings $g_B$ and $g_I$ are specified as follows
\begin{eqnarray}
g^{-1}_B=-\frac{\Gamma({{2-D}\over 2})}{(4\pi)^{D\over 2}}\left(\frac{m}{\hbar^2}\right)^{D\over 2}|\epsilon_B|^{{D\over 2}-1},\label{g_phys_B}\\
g^{-1}_I=-\frac{\Gamma({{2-D}\over 2})}{(2\pi)^{D\over 2}}\left(\frac{m}{\hbar^2}\right)^{D\over 2}|\epsilon_I|^{{D\over 2}-1},\label{g_phys_I}
\end{eqnarray}
where $\Gamma(z)$ stands for the gamma function. Note that the bound states are only possible for positive $g_B$s and $g_I$s, but it is convenient to parametrize negative couplings by the binding energies either. By careful inspection of the $D\to 2$ limit one can conclude that Eqs.~(\ref{g_bare_B}), (\ref{g_bare_I}) and (\ref{g_phys_B}), (\ref{g_phys_I}) provide a correct description of zero-range potentials even in the two-dimensional case. Moreover, the $D=2$ pseudo-potential always provides the existence of one bound state.

The alternative way (see, for instance \cite{Volosniev2015}) to deal with a point-like interactions is to initially start from some `physical` (Gaussian, for instance) potentials and then relate the appropriate coupling constant to the $s$-wave scattering lengths $a_B$ and $a_I$ in the limit where the effective ranges are the smallest parameters with the dimension of length in the system. In the following, no restrictions are set on a magnitude of the boson-impurity interaction, while the Bose gas itself is expected to be extremely dilute.

\subsection{Effective field theory approach}
The further analysis will be performed in a spirit of the effective field theory approach (see, for review \cite{Andersen2004}), which is known to be extremely convenient for the many-boson systems. Particularly, this formulation automatically guarantees the implementation of the Hugengoltz-Pines theorem (which is a concrete manifestation of the Goldstone theorem) in every order of a loop expansion. Moreover, the effective field theory approach provides a non-perturbative predictions for the Bose gas thermodynamics. In the limit of weak boson-boson coupling the loop expansion is identical to the perturbation theory in term of characteristic small parameter $a^D_Bn$. The main idea of the method relies on the separation of `classical' dynamics during the computations of partition function by means of the path integral
\begin{eqnarray}\label{psi}
\psi(x)=\psi_0({\bf r})+\tilde{\psi}(x),\ \
\psi^*(x)=\psi^*_0({\bf r})+\tilde{\psi}^*(x),
\end{eqnarray}
where the introduced classical fields are determined by the minimization of the action (\ref{S}): $\delta S_0=\delta S[\psi^*_0,\psi_0]=0$. Note that in general $|\psi_0({\bf r})|^2$ should not be confused with the Bose condensate density. In the absence of impurities, $\Phi({\bf r})=0$, the solution $\psi_0({\bf r})$ is real and uniform. Putting a microscopic amount of heavy particles in the Bose condensate we cannot principally change the character of this solution provided that $\psi_0({\bf r})$ becomes only slightly non-uniform, i.e., $\int_{L^D}d{\bf r}|\psi_0({\bf r})|^2\propto L^D$. Of course, one may argue that the localized solutions $\psi_0({\bf r})$ decrease the total energy by $\propto-\mathcal{N}|\epsilon_I|$, but any non-zero repulsion between bosons immediately increases the energy of the system by $\propto N^2g_B/a^D_I$. Therefore, the collapsed BEC state \cite{Panochko2021} is not energetically preferable in the thermodynamic limit, where both number of the repulsively interacting bosons $N$ and volume of the box $L^D$ infinitely increase.

Performing the shift (\ref{psi}), we end up with the following effective action
\begin{eqnarray}\label{S_eff}
S_{\textrm{eff}}=S_0-\frac{1}{2}\int d x\left[\tilde{\psi}^*(x),\tilde{\psi}(x)\right]\hat{K}\left[ \begin{array}{c}
\tilde{\psi}(x)\\
\tilde{\psi}^*(x)\\
\end{array} \right],
\end{eqnarray}
where only the Gaussian in the fluctuation fields part is explicitly written down. Here the $2\times 2$ matrix operator $\hat{K}$ with elements
\begin{eqnarray}\label{K}
&&\hat{K}_{11}=\varepsilon-\mu+\Phi({\bf r})+2g_{B,\Lambda}|\psi_0({\bf r})|^2-\partial_{\tau},\nonumber\\
&&\hat{K}_{12}=\hat{K}^*_{21}=g_{B,\Lambda}\psi^2_0({\bf r}),\nonumber\\
&&\hat{K}_{22}=\varepsilon-\mu+\Phi({\bf r})+2g_{B,\Lambda}|\psi_0({\bf r})|^2+\partial_{\tau}.
\end{eqnarray}
is introduced. Taking into account the equation for $\psi_0({\bf r})$
\begin{eqnarray}\label{Eq_psi_0}
\left\{\varepsilon-\mu+\Phi({\bf r})+g_{B,\Lambda}|\psi_0({\bf r})|^2\right\}\psi_0({\bf r})=0,
\end{eqnarray}
and performing the Gaussian integration in (\ref{S_eff}), we finally obtain the grand potential of the Bose system with the impurities immersed
\begin{eqnarray}\label{Omega}
\Omega = -\frac{g_{B,\Lambda}}{2}\int_{L^D}d{\bf r}|\psi_0({\bf r})|^4+\frac{1}{2\beta}\textrm{Sp}\ln\hat{K}-\textrm{const},
\end{eqnarray}
where $\textrm{Sp}$ denotes the trace in the $D+1$ space. A constant term (counterterm) in (\ref{Omega}), is most straightforwardly represented in the plane-wave basis $\textrm{const}=\frac{1}{2}\sum_{\bf k}\langle {\bf k}|\varepsilon-\mu+2g_{B,\Lambda}|\psi_0({\bf r})|^2+\Phi({\bf r})|{\bf k}\rangle$, but cannot be obtained by the functional integration and has to be written by hands \cite{Salasnich2016} in order to resolve a standard normal-ordering routine. Consequently, the calculation of thermodynamics for `Bose gas + static impurities' reduces to finding a solution of Eq.~(\ref{Eq_psi_0}), and then with $\psi_0({\bf r})$ in hands to the evaluation of the functional determinant. Note that by taking into account $S_0$ only, one reproduces the mean-field \cite{Volosniev2017,Pastukhov2019,Panochko2019,Hryhorchak2020,Hryhorchak2020_2,Massignan2021} description of the system generalized to $\mathcal{N}$ impurities in the static limit.

\subsection{Limit of dilute Bose gas}
In general case, the above program, which can be realized to the very end in 1D \cite{Reichert2019} even at finite impurity masses \cite{Volosniev2017,Panochko2019,Jager2020}, requires considerable numerical efforts in the higher dimensions, but the limit of weak inter-boson interaction can be handled more or less easily. Indeed, the intrinsic, for the dilute Bose gas, length-scale is represented by the so-called coherence length $\xi=\frac{\hbar}{mc}$ (with $c=\sqrt{ng_B/m}$ being the sound velocity), which is large in comparison to the average distance between particles and to the $s$-wave scattering length $a_B$. The magnitude of boson-impurity interaction, in turn, is dictated by $a_I$. So if we additionally assume that $a_I\ll \xi$, the solution of Eq.~(\ref{Eq_psi_0}) can be immediately found $\psi_0({\bf r})=\sqrt{\mu/g_{B,\Lambda}}\simeq \sqrt{n}$. In all other cases, we can apply the successive expansion in terms of the $\psi_0$-field `non-uniformity'
\begin{eqnarray}\label{psi_0_solution}
\psi_0({\bf r})=\sqrt{\mu/g_{B,\Lambda}}\left\{1-\bar{\psi}^{(1)}_0({\bf r})-\bar{\psi}^{(2)}_0({\bf r})\ldots \right\},
\end{eqnarray}
where after the substitution in Eq.~(\ref{Eq_psi_0}) the dimensionless functions $\bar{\psi}^{(1)}_0({\bf r})$, $\bar{\psi}^{(2)}_0({\bf r})$ satisfy the following equations:
\begin{eqnarray}
&&\left\{\varepsilon+2\mu+\Phi({\bf r})\right\}\bar{\psi}^{(1)}_0({\bf r})=\Phi({\bf r}),\label{barpsi^1_0}\\
&&\left\{\varepsilon+2\mu+\Phi({\bf r})\right\}\bar{\psi}^{(2)}_0({\bf r})=3\mu \left(\bar{\psi}^{(1)}_0({\bf r})\right)^2.\label{barpsi^2_0}
\end{eqnarray}
Note that the above approximate procedure does not require the boson-impurity interaction to be weak. Furthermore, by a naive dimensional analysis, it is easy to argue that both at the weak and strong couplings $g_I$, the contribution of the second-order correction $\bar{\psi}^{(2)}_0({\bf r})$ in the thermodynamics of the system is much smaller than the one originating from $\bar{\psi}^{(1)}_0({\bf r})$. Therefore, in our consideration below we fully focus on the first-order correction. But even this simple approximation effectively sums up some infinite set of terms of the standard pertubation theory for a model with the uniform condensate \cite{Kain2018}. Equation (\ref{barpsi^1_0}) with $\Phi({\bf r})$ given by (\ref{Phi}) can be solved for arbitrary $\mathcal{N}$ by means of the Fourier transformation
\begin{eqnarray}\label{barpsi^1_0_sol}
\bar{\psi}^{(1)}_0({\bf r})=\sum_{1\le j\le \mathcal{N}}A_j\frac{1}{L^D}\sum_{\bf k}\frac{e^{{\rm i}{\bf k}({\bf r}-{\bf r}_j)}}{\varepsilon_k+2\mu},
\end{eqnarray}
with $\varepsilon_k=\frac{\hbar^2k^2}{2m}$ and coefficients $A_j=\sum_{1\le i\le \mathcal{N}}T_{ji}(-2\mu)$, where matrix $T_{ji}(-2\mu)$ is introduced in Appendix.

We can now proceed with the calculations of the functional determinant in (\ref{Omega}). Taking into account the extreme diluteness of the Bose subsystem, it is enough to expand $\textrm{Sp}\ln\hat{K}\simeq\textrm{Sp}\ln\hat{K}^{(0)}+\textrm{Sp}\left\{[\hat{K}^{(0)}]^{-1}\Delta\hat{K}\right\}$, where $\hat{K}^{(0)}$ is given by (\ref{K}) but with $\psi_0({\bf r})\to \sqrt{\mu/g_{B,\Lambda}}$ and $\Delta\hat{K}=\hat{K}-\hat{K}^{(0)}$. Following our previous discussion, we ignore in $\Delta\hat{K}$ all higher-order corrections except $\bar{\psi}^{(1)}_0({\bf r})$. After this, the calculations are relatively simple and at absolute zero we obtain the $\Omega$-potential in the adopted approximation
\begin{eqnarray}\label{Omega_approx}
&&\Omega \simeq - L^D\frac{\mu^2}{2g_{B,\Lambda}}+\frac{\mu}{g_{B,\Lambda}}\sum_{1\le j\le \mathcal{N}}A_j\nonumber\\
&&+\frac{1}{2}\sum_{\bf k}\langle {\bf k}|\mathcal{E}-\varepsilon-\mu-\Phi({\bf r})|{\bf k}\rangle\nonumber\\
&&+\frac{1}{L^D}\sum_{\bf k}\left\{1-\frac{\varepsilon_k+\mu/2}{E_k}\right\}\sum_{1\le j\le \mathcal{N}}A_j,
\end{eqnarray}
where $\mathcal{E}=\sqrt{(\varepsilon+\Phi({\bf r}))^2+2\mu(\varepsilon+\Phi({\bf r}))}$ and $E_k=\sqrt{\varepsilon_k^2+2\mu\varepsilon_k}$ stands for the Bogoliubov spectrum of the `pure' Bose system. It should be noted that for dilute Bose systems the impact of the so-called quantum fluctuations (terms with the summations over the wave-vector) to $\Omega$ is much smaller than the first two terms (the mean-field contributions). The last step to be performed in these calculations is to replace the bare couplings $g_{B,\Lambda}$ and $g_{I,\Lambda}$ via (\ref{g_bare_B}) and (\ref{g_bare_I}), respectively. This procedure provides the convergence of sums over the wave-vector in last two terms of (\ref{Omega_approx}). Then the trace in the third term of (\ref{Omega_approx}) can be computed (see, Appendix for details). With the well-defined grand potential, we can relate, by using the thermodynamic identity $n=-\frac{\partial}{\partial \mu}\frac{\Omega}{L^D}$, the chemical potential of the Bose system to its equilibrium density $n$. Performing these calculations, one must keep in mind that the presence of a microscopic number of impurities cannot principally change the properties of the system. So, if we denote (and appropriate grand potential $\Omega_B$) the chemical potential of Bose gas without exterior particles by $\mu_B$, the difference $\Delta\mu=\mu-\mu_B\propto \mathcal{N}/L^D$ should be small. Using this fact and $n=-\frac{\partial}{\partial \mu}\frac{\Omega_B}{L^D}-\frac{\partial}{\partial \mu}\frac{\Delta\Omega}{L^D}$, we can identify a small correction $\Delta\mu=-\frac{\partial \Delta\Omega}{\partial \mu_B}/\frac{\partial^2 \Omega_B}{\partial \mu_B^2}$. The latter formula allows to determine the energy that the Bose system gains when $\mathcal{N}$ impurities are immersed
\begin{eqnarray}
\Delta E_{\mathcal{N}}=\left(\Omega-\Omega_B\right)|_{\mu\to \mu_B},
\end{eqnarray}
which is an explicit manifestation of the well-known theorem about small corrections to the thermodynamic potentials.

\section{Results}
Before we proceed to describing our main results, it is necessary to analyze the case of `pure' bosons. Setting $\Phi({\bf r})=0$ in (\ref{Omega_approx}) and calculating integrals, we obtain for density
\begin{eqnarray}
n=\frac{\mu_B}{g_{B}}\left\{1-\frac{\Gamma(D)}{{D\over 2}\Gamma^2({D\over 2})}\left(\frac{\mu_B}{|\epsilon_B|}\right)^{{D\over 2}-1}\right\},
\end{eqnarray}
which allows to obtain expression for $\mu_B$ iteratively. For the weakly non-ideal three-dimensional bosons we find the well-known formula ($|\epsilon_B|=\frac{\hbar^2}{ma_{B}^2}$ in 3D)
\begin{eqnarray}
\mu_B=\frac{4\pi\hbar^2a_{B}n}{m}\left\{1+\frac{32}{3\sqrt{\pi}}\sqrt{na_{B}^3}+\ldots\right\}.
\end{eqnarray}
Similarly, in the two-dimensional case we have the transcendental equation \cite{Mora2009}
\begin{eqnarray}
n=\frac{m\mu_B}{4\pi\hbar^2}\left\{\ln\frac{|\epsilon_B|}{\mu_B}-1\right\}.
\end{eqnarray}
Being convinced that the limit of Bose gas without impurities is correctly reproduced by the adopted approach, we are ready to present our main results concerning the binding energy of one and two impurity atoms in the dilute three- and two-dimensional Bose gases.

\subsection{3D case}
\begin{figure}[h!]
	\includegraphics[width=0.35\textwidth,clip,angle=-0]{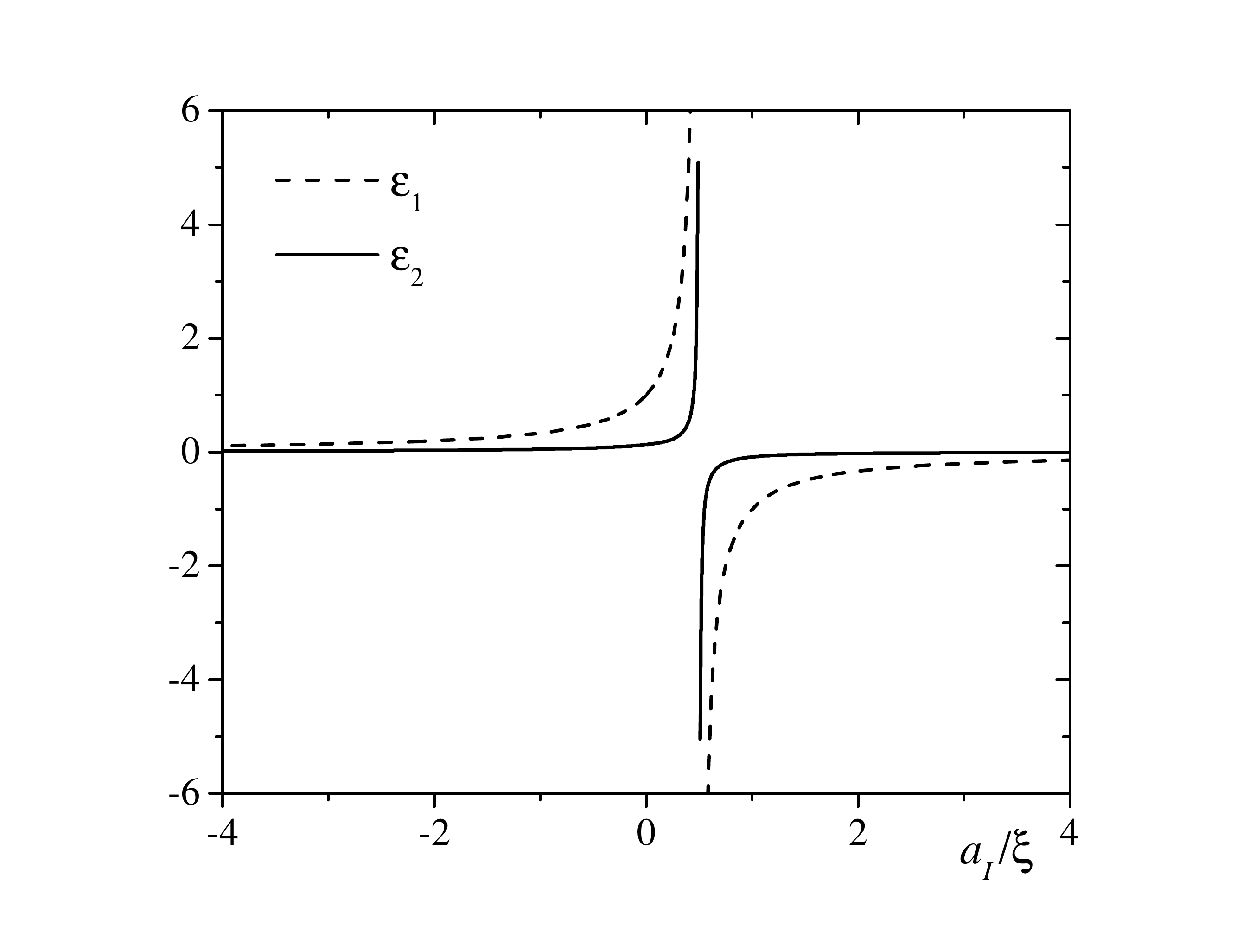}
	\caption{Dimensionless functions $\varepsilon_{1,2}\left(\frac{a_I}{\xi};\infty\right)$ determining the one-impurity energy in 3D dilute Bose gas.}\label{one_particle_3D}
\end{figure}

\begin{figure}[h!]
	\includegraphics[width=0.35\textwidth,clip,angle=-0]{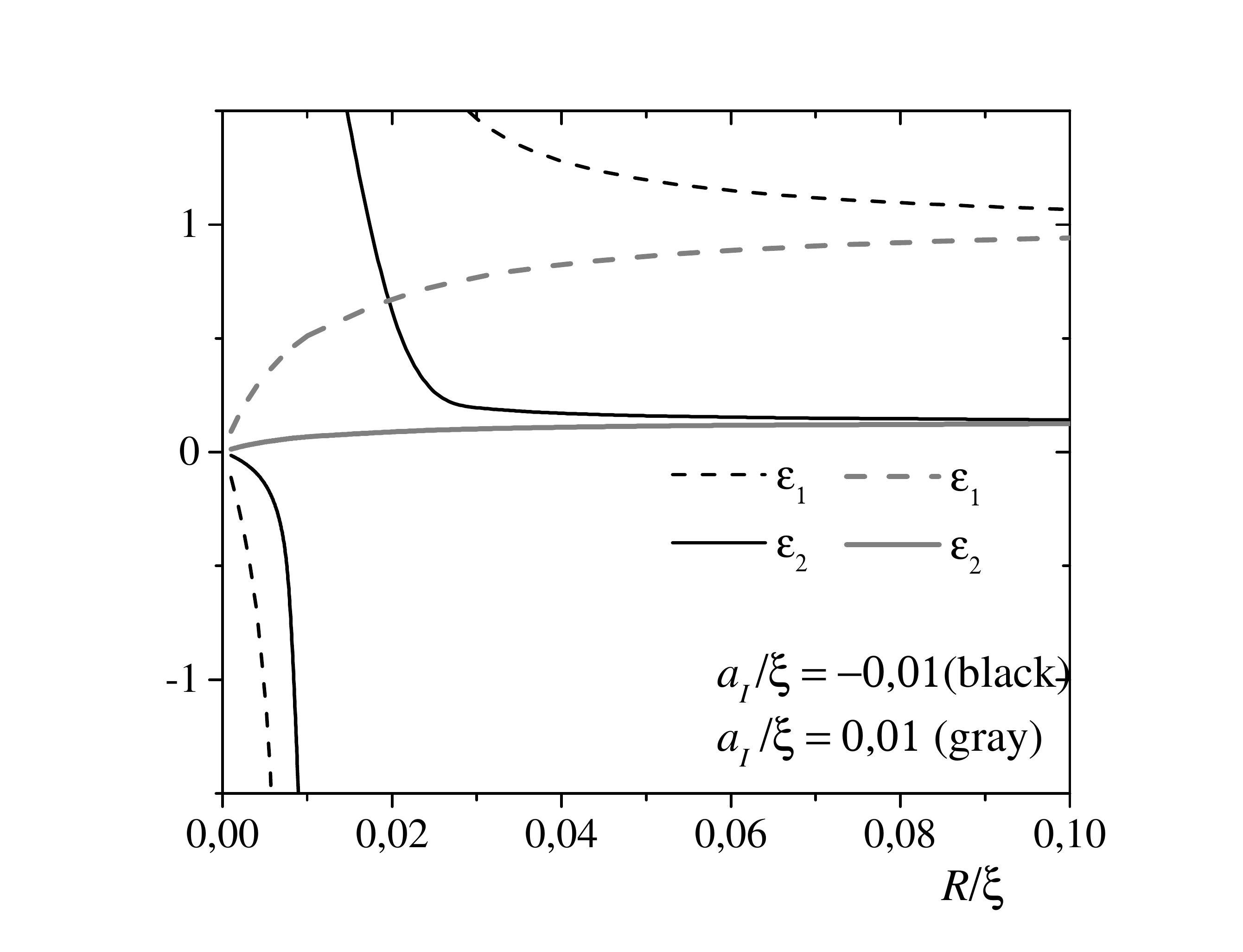}
	\includegraphics[width=0.35\textwidth,clip,angle=-0]{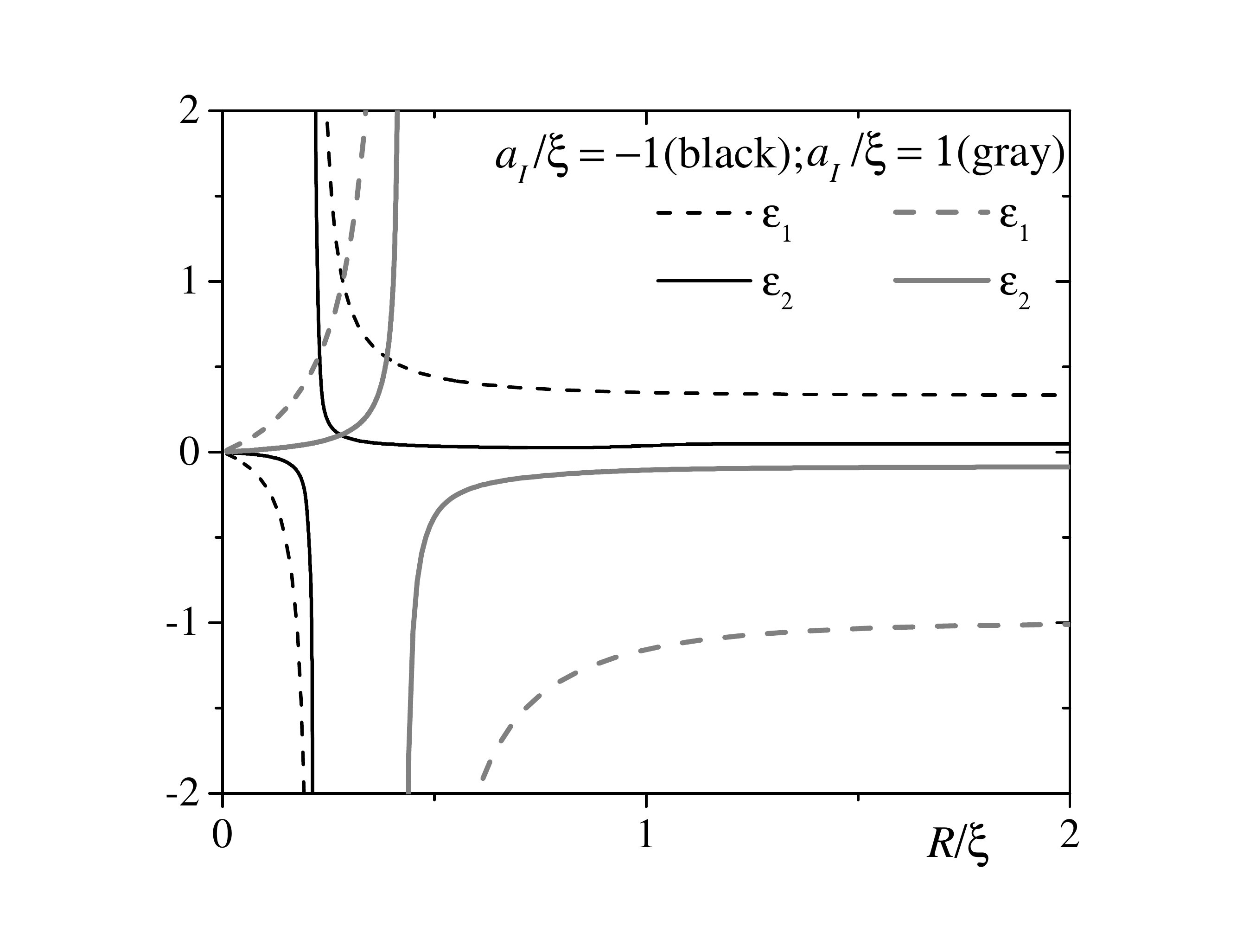}
	\caption{Mean-field and the first-order quantum corrections $\varepsilon_{1,2}\left(\frac{a_I}{\xi};\frac{R}{\xi}\right)$ to the energy of 3D dilute Bose gas generated by two impurities for $\frac{a_I}{\xi}=\pm 0.01$ and $\frac{a_I}{\xi}=\pm 1$.}\label{two_particle_3D}
\end{figure}

In 3D, the general structure of the two-impurity binding energy in the dilute Bose gas $(n\xi^3\ll 1)$ can be represented as
\begin{eqnarray}\label{E_2_3D}
\Delta E_2= \Delta E^{(0)}_2\left[\varepsilon_1\left(\frac{a_I}{\xi};\frac{R}{\xi}\right)+\frac{1}{n\xi^3}\varepsilon_2\left(\frac{a_I}{\xi};\frac{R}{\xi}\right)+\ldots\right],
\end{eqnarray}
where $\Delta E^{(0)}_2=2g_In$ is the contribution of the ideal Bose gas, $a_I$ is the $s$-wave scattering length that parametrizes the (renormalized) two-body coupling $g_I=\frac{2\pi\hbar^2a_I}{m}$ and $R$ is the distance between two static particles. The first term in (\ref{E_2_3D}) has a simple analytic form
\begin{eqnarray}
\varepsilon_1\left(\frac{a_I}{\xi};\frac{R}{\xi}\right)=\frac{\xi/a_I}{\xi/a_I-2+e^{-2R/\xi}/(R/\xi)},
\end{eqnarray}
and originates purely from the mean-field correction to the grand potential [the second term in (\ref{Omega_approx})], while $\varepsilon_2\left(\frac{a_I}{\xi};\frac{R}{\xi}\right)$ contains both the mean-field and purely quantum corrections. Note that in formula for $\Omega$ only the one-loop corrections were taken into account and a consistent consideration of the next to a leading order terms in series expansion over the small parameter $1/(n\xi^3)$ necessary requires the calculation of the two-loop diagrams to the grand potential. By setting the distance between heavy particles $R$ to infinity, one obtains from (\ref{E_2_3D}) the one-impurity limit. A typical behavior of functions $\varepsilon_{1,2}\left(\frac{a_I}{\xi};\infty\right)$ is presented in Fig.~\ref{one_particle_3D}.
Let us recall that the problem considered here is the exactly solvable one, when the bosons are non-interacting. Therefore, it should be clearly understood that the presented results are accurate if the coherence length $\xi$ is the largest parameter with dimension of length in the system. In order to reveal the interplay between regimes of very dilute $a_I/\xi \to 0$ Bose gas and intermediate boson-impurity interaction we have plotted in Fig.~\ref{two_particle_3D} the binding energy of two heavy particles for the positive and negative $s$-wave scattering lengths $a_I$.
Comparing these findings to the ideal Bose gas results \cite{Panochko2021}, we can observe similar patterns in the behavior of the systems at weak coupling: at positive $a_I$ the binding energy is the monotonic function of $R$, while at the negative boson-impurity scattering lengths both $\varepsilon_{1,2}\left(-0.01;\frac{R}{\xi}\right)$ have a simple-pole singularity. When the interaction increases (see lower panel in Fig.~\ref{two_particle_3D}) the mean-field and quantum corrections to the ground state energy of 3D Bose gas possess an infinite discontinuities independently of a sign of $a_I$.

\subsection{2D case}
\begin{figure}[h!]
	\includegraphics[width=0.35\textwidth,clip,angle=-0]{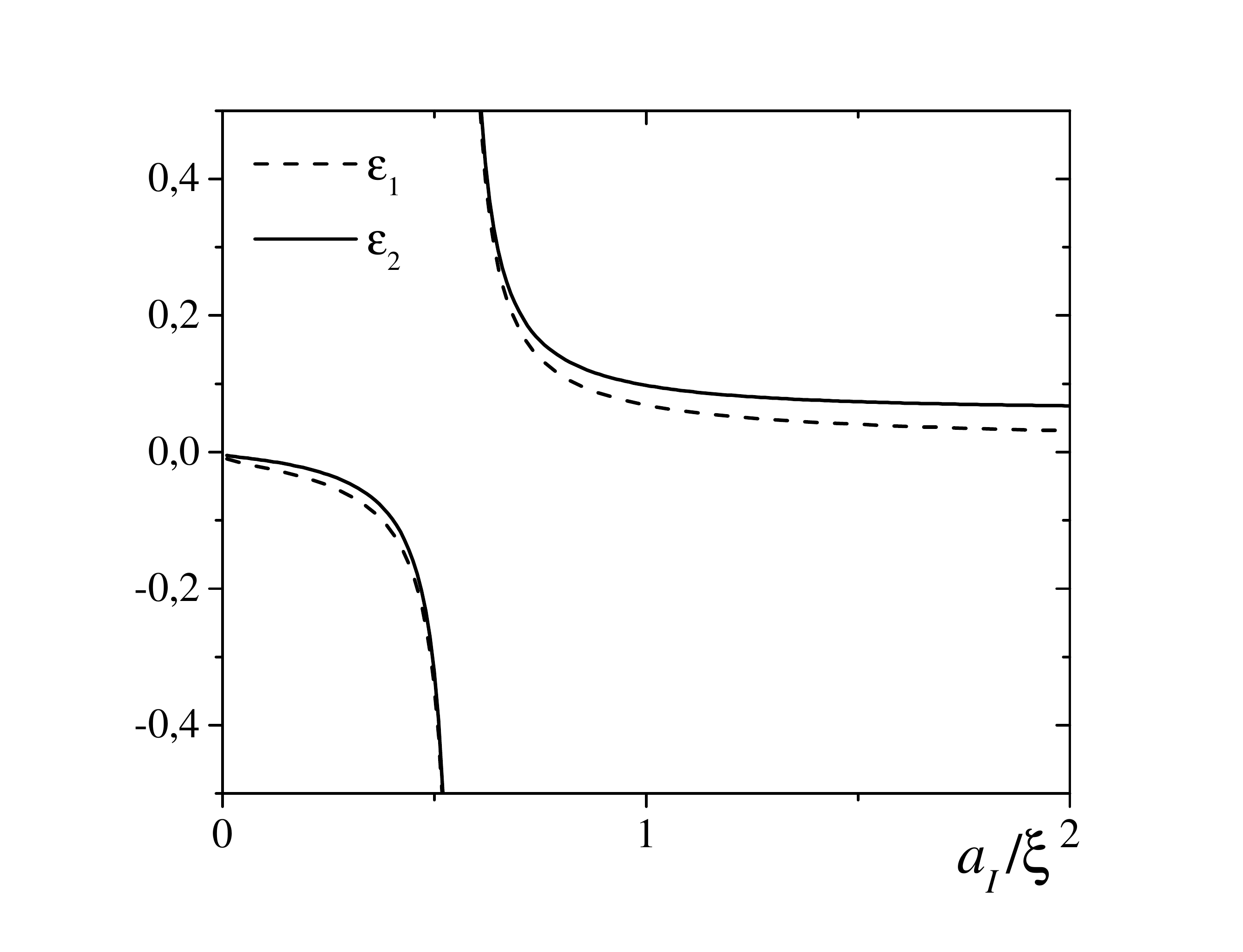}
	\caption{One-impurity binding energy $\varepsilon_{1,2}\left(\frac{a_I}{\xi};\infty\right)$ (see Eq.~(\ref{E_2_2D})) in 2D case.}\label{one_particle_2D}
\end{figure}

In general, the low-dimensional dilute Bose systems with static impurities are very peculiar. When the interaction between bosons is switched off, these systems are insensible to the boson-impurity interaction in their not collapsed ground state, and therefore, the binding energy of the heavy particles requires a finite compressibility of the host system to be non-zero. Introducing the two-body $s$-wave scattering length $a_I$ through the boson-impurity vacuum bound state energy $|\epsilon_I|=2e^{-2\gamma}\hbar^2/(ma^2_I)$, we can write down the energy that the 2D Bose gas gains when two heavy particles are immersed in it
\begin{eqnarray}\label{E_2_2D}
\Delta E_2= 2\frac{2\pi\hbar^2n}{m}\left[\varepsilon_1\left(\frac{a_I}{\xi};\frac{R}{\xi}\right)\right.\nonumber\\
\left.+\frac{1}{n\xi^2}\varepsilon_2\left(\frac{a_I}{\xi};\frac{R}{\xi}\right)+\ldots\right].
\end{eqnarray}
Note that in contrast to a 3D case, both $\varepsilon_{1,2}\left(\frac{a_I}{\xi};\frac{R}{\xi}\right)$ tend to zero (at least logarithmically) in the limit of ideal Bose gas ($\xi \to \infty$). At large distances $R$, Eq.~(\ref{E_2_2D}) gives the double binding energy of a single impurity which is presented in Fig.~\ref{one_particle_2D}.
Particularly, these calculations clearly demonstrate the weakening of the role of quantum fluctuations in the formation of polarons in two-dimensional Bose systems. Actually, this observation \cite{Jager2020} seems to be intrinsic for the low-dimensional systems in general.

\begin{figure}[h!]
	\includegraphics[width=0.35\textwidth,clip,angle=-0]{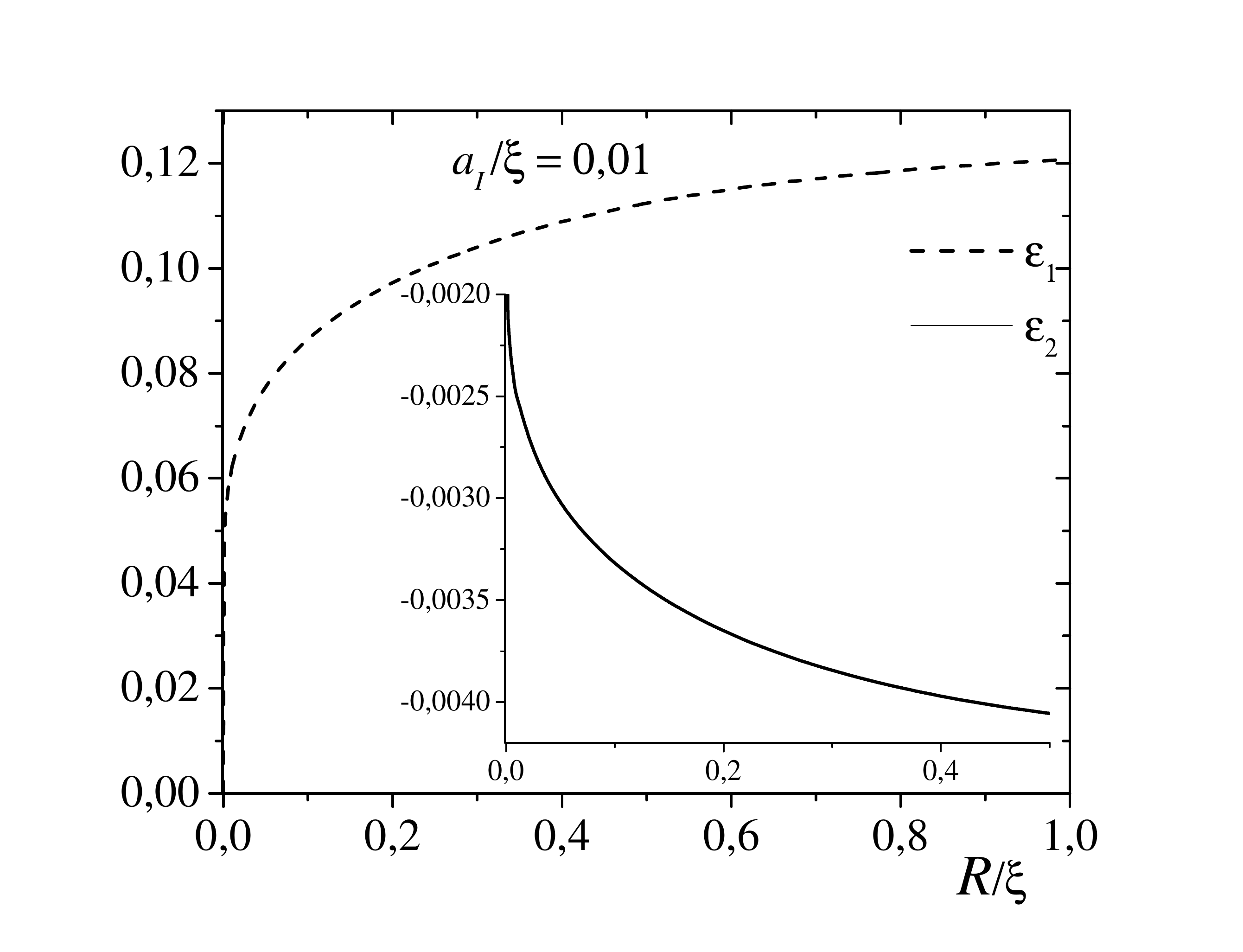}
	\includegraphics[width=0.35\textwidth,clip,angle=-0]{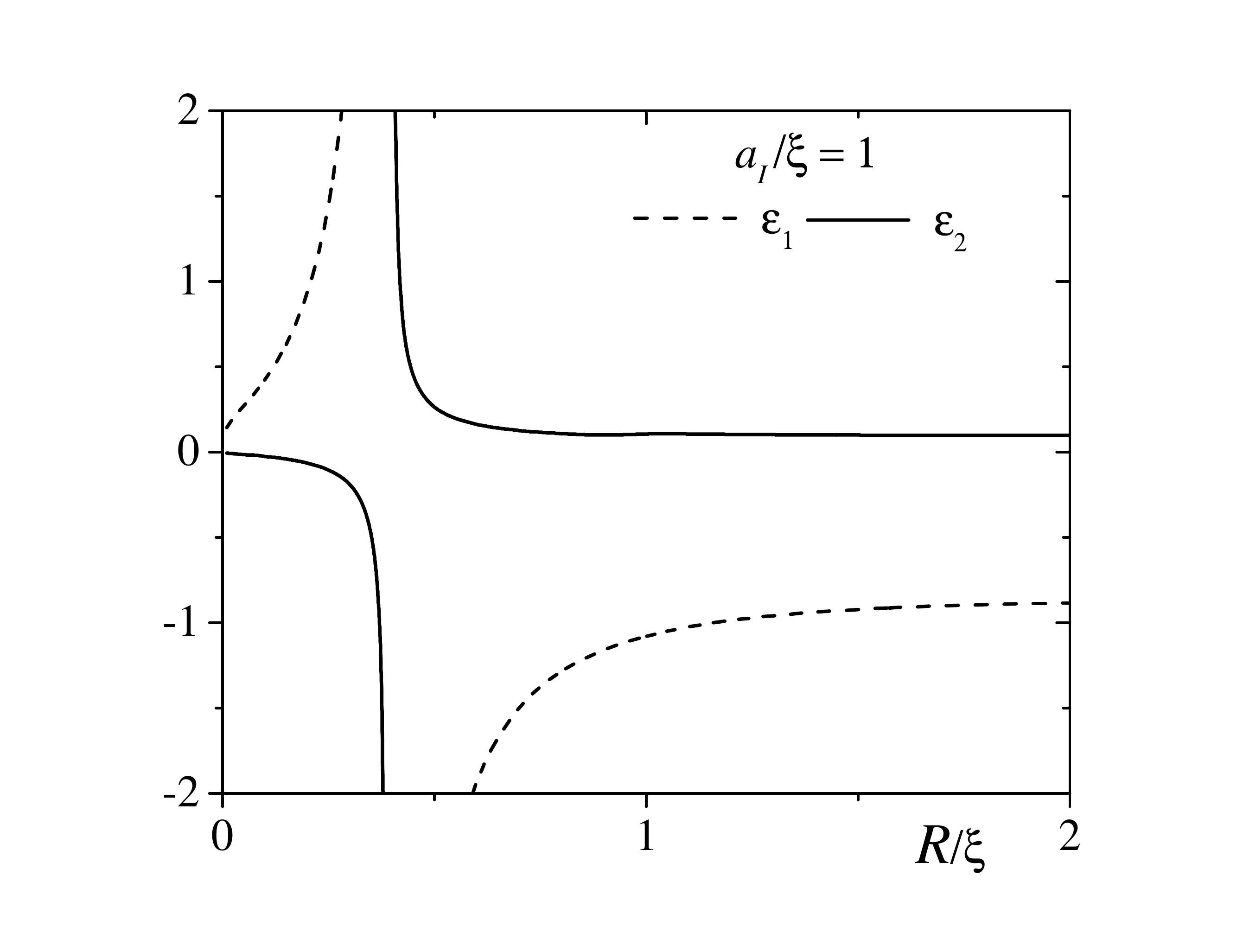}
	\caption{Two-impurity binding energies $\varepsilon_{1,2}\left(\frac{a_I}{\xi};\frac{R}{\xi}\right)$ in 2D dilute Bose gas.}\label{two_particle_2D}
\end{figure}

The numerical computations of the two-impurity energies (see Fig.~\ref{two_particle_2D})
in the 2D Bose gas demonstrate qualitative similarity between the two- and three-dimensional cases. At weak boson-impurity interactions $a_I/\xi \ll 1$, where our effective field-theoretical formulation is supposed to make a quantitative predictions, the mean-field term $\varepsilon_{1}\left(\frac{a_I}{\xi};\frac{R}{\xi}\right)$ as well as the one that includes the quantum corrections $\varepsilon_{2}\left(\frac{a_I}{\xi};\frac{R}{\xi}\right)$ behave as monotonic functions of $R$. The interaction-induced effective two-body potential between static particles at large $a_I/\xi$ always contains singularity.

\section{Conclusions}
In summary, by means of the effective field theory formulation, we have calculated the impurity-induced shifts to the ground state energies of the two- and three-dimensional dilute Bose gases. Particularly, by taking into account the extreme diluteness of the host bosons, we have proposed the approximate procedure that allows to calculate the properties of an arbitrary (microscopic) number of static impurities in terms of characteristic small parameter $1/(n\xi^D)$ (where $n$ and $\xi$ are the density and coherence length of bosons, respectively). The numerical calculations of the binding energies of two static impurities in dilute 2D and 3D Bose gases that were performed for a wide range of the boson-impurity interactions and distances between impurities has revealed the peculiarities of the medium-induced (Casimir) forces: \textit{i)} the two-body effective potential always demonstrates singular behavior at the distances between impurities comparable to the boson-impurity $s$-wave scattering lengths $a_I$; \textit{ii)} an impact of purely quantum corrections decreases with the lowering of a spatial dimensionality. Similar singularities are also intrinsic for the binding energy of a single impurity at $a_I\sim \xi$, which may signal \cite{Schmidt2021} about the inapplicability of the adopted approximate treatment for calculations of the `classical' solution $\psi_0({\bf r})$ in that region.

\section{Appendix}
For completeness, in this section we give some details of the calculations not presented in the main text. Let us first start from the equation that determines the classical field $\psi_0({\bf r})$. Explicitly writing down Eq.~(\ref{barpsi^1_0}), after the implementation of ansatz (\ref{barpsi^1_0_sol})
\begin{eqnarray*}
\sum_{1\le j\le \mathcal{N}}A_j\delta_{\Lambda}({\bf r}-{\bf r}_j)+\sum_{1\le j\le \mathcal{N}}g_{I,\Lambda}\delta_{\Lambda}({\bf r}-{\bf r}_j)\\
\times\sum_{1\le i\le \mathcal{N}}A_i\frac{1}{L^D}\sum_{\bf k}\frac{e^{{\rm i}{\bf k}({\bf r}-{\bf r}_i)}}{\varepsilon_k+2\mu}=\sum_{1\le j\le \mathcal{N}}g_{I,\Lambda}\delta_{\Lambda}({\bf r}-{\bf r}_j),
\end{eqnarray*}
and combining $j=i$ terms in double sum with the first term of equation, we obtain
\begin{eqnarray*}
	A_j\left[\frac{1}{g_{I,\Lambda}}+\frac{1}{L^D}\sum_{\bf k}\frac{1}{\varepsilon_k+2\mu}\right]+\\
	\sum_{1\le i\neq j\le \mathcal{N}}\frac{1}{L^D}\sum_{\bf k}\frac{e^{{\rm i}{\bf k}({\bf r}_j-{\bf r}_i)}}{\varepsilon_k+2\mu}A_i=1.
\end{eqnarray*}
The divergent sum in the square brackets is now regularized by the renormalization of a coupling constant (\ref{g_bare_I}), so the final result contains only observable $g_I$. One can easily recognize the square brackets as the boson-impurity two-body $T$-matrix
\begin{eqnarray*}
	t^{-1}_I(\omega)=g^{-1}_{I,\Lambda}-\frac{1}{L^D}\sum_{\bf k}\frac{1}{\omega-\varepsilon_k},
\end{eqnarray*}
and introducing auxiliary notations
\begin{eqnarray*}
\Delta_{ij}(\omega)=\frac{1}{L^D}\sum_{\bf k}\frac{e^{{\rm i}{\bf k}({\bf r}_i-{\bf r}_j)}}{\omega-\varepsilon_k},
\end{eqnarray*}
we find the result for coefficients $A_j$ announced in the main text
\begin{eqnarray*}
A_i=\sum_{1\le j\le \mathcal{N}}T_{ij}(-2\mu),\\
T^{-1}_{ij}(-2\mu)=\delta_{ij}t^{-1}_I(-2\mu)-\Delta_{ij}(-2\mu)(1-\delta_{ij}).
\end{eqnarray*}

For the calculation of trace in the second term of (\ref{Omega_approx}), we have used formal identity
\begin{eqnarray*}
\sum_{\bf k}\langle {\bf k}|\mathcal{E}-\varepsilon-\mu-\Phi({\bf r})|{\bf k}\rangle=\\
\int d\omega D(\omega)\left[\sqrt{\omega^2+2\mu \omega}-\omega-\mu\right],\\
D(\omega)=\sum_{\bf k}\langle {\bf k}|\delta(\omega-\varepsilon-\Phi({\bf r})|{\bf k}\rangle.
\end{eqnarray*}
The density of states $D(\omega)$ is easily calculated within the Green's function method \cite{Panochko2021}
\begin{eqnarray*}
	D(\omega)=\sum_{\bf k}\left[\delta(\omega-\varepsilon_k)-\frac{1}{\pi}{\rm Im} \frac{\langle {\bf k}|\mathcal{T}(\omega+{\rm i}0)|{\bf k}\rangle}{(\omega+{\rm i}0-\varepsilon_k)^2}\right],
\end{eqnarray*}
where the $T$-matrix $\mathcal{T}(\omega)$ characterizes the scattering of a single boson on $\mathcal{N}$ impurities
\begin{eqnarray*}
\langle {\bf q}|\mathcal{T}(\omega)|{\bf k}\rangle=\sum_{1\le i,j\le \mathcal{N}}e^{-{\rm i}{\bf q}{\bf r}_i}T_{ij}(\omega)e^{{\rm i}{\bf k}{\bf r}_j}.
\end{eqnarray*}

The calculations of $\langle {\bf k}|\mathcal{T}(\omega+{\rm i}0)|{\bf k}\rangle$ in the density of states requires the knowledge of an explicit analytic formulas for the boson-impurity two-body $T$-matrix
\begin{eqnarray*}
	t^{-1}_{I}(\omega)=\frac{\Gamma({{2-D}\over 2})}{(2\pi)^{D\over 2}}\left(\frac{m}{\hbar^2}\right)^{D\over 2}\left[(-\omega)^{{D\over 2}-1}-|\epsilon_I|^{{D\over 2}-1}\right],
\end{eqnarray*}	
and a function $\Delta_{ij}(\omega)=\Delta_{R}({\omega})$ of distance $R=|{\bf r}_i-{\bf r}_j|$ between two impurities in arbitrary $D$
\begin{eqnarray*}
	 \Delta_{R}(\omega)=\frac{1}{(2\pi)^{D\over 2}}\frac{2mk^{D-2}_{\omega}}{\hbar^2}\frac{K_{{D\over 2}-1}(Rk_{\omega})}{\left(Rk_{\omega}\right)^{{D\over 2}-1}},
\end{eqnarray*}
where $k_{\omega}=\sqrt{2m(-\omega)}/\hbar$, and $K_{\nu}(z)$ is the modified Bessel function of the second kind \cite{Abramowitz}.

\begin{center}
	{\bf Acknowledgements}
\end{center}
We are indebted to Dr.~Iryna Pastukhova for comments on the manuscript.

\end{document}